\documentclass[preprint,12pt]{article}
\usepackage{authblk}
\usepackage{amsmath, amssymb, amsthm}
\usepackage[margin=1in]{geometry}

\usepackage{tikz}
\usepackage{tikz-cd}
\usepackage{booktabs}
\usepackage{tabularx}
\usepackage{longtable}
\usetikzlibrary{cd}

\usepackage{enumitem}
\usepackage{array}
\usepackage{float} 

\newtheorem{definition}{Definition}

\theoremstyle{remark}
\newtheorem{remark}{Remark}
\newtheorem{example}{Example}

\begin{document}

\title{Construction Defining Functionality:\\ 
A Constructive Perspective on Functions\\ 
through Their Generated Structures}

\author{Yumiko Nishiyama}
\date{Octber,2025}

\maketitle

\begin{abstract}
In this work, we propose the concept of Construction Defining Functionality (CDF), which characterizes functions by the structural spaces they generate through iteration, recursion, and logical application. By viewing functions as generators of hierarchical structures, we formalize these generated structural spaces and organize a framework to classify and mathematically model their properties. The organized CDF framework captures the intrinsic constructive behaviors of functions via their generated structural spaces.
\end{abstract}

\section{Introduction}

In this work, we define the constructive properties and behaviors of functions as \textbf{Construction Defining Functionality (CDF)} through their generated spaces. This concept reflects a perspective in which mathematical functions are reinterpreted as \textit{spatial constructors} or \textit{structure generators}. This framework serves not only as a tool for classifying and analyzing functions based on their generative complexity, but also as a conceptual lens through which deeper mathematical structures can be examined and understood. By emphasizing the spaces that functions generate, CDF facilitates a structural understanding that transcends traditional input–output characterization, potentially enriching various areas of mathematics and logic.

\section{Conceptualizing Generated Structural Spaces}

Given a function \( f: X \to Y \), we examine how it generates structures through iteration, recursion, or functional application.

This motivates the following abstract inquiries:

\begin{enumerate}
  \item \textbf{Conceptualization of Generated Structural Spaces:}  
  What kind of structures can functions generate? How can these be defined, classified, and compared formally?

  \item \textbf{Mathematical Modeling of Hierarchical Systems:}  
  Can these structural classifications be formalized as mathematical models or frameworks for representing hierarchical spaces?
\end{enumerate}

These considerations form the foundation for the formal definition of the structural space \( S(f) \), and the development of the CDF framework.

\section{Definition of CDF}

\textbf{Definition 1 (Generated Structural Space via CDF).}  
Given a function \( f : X \to Y \), the structural space \( S(f) \) generated by \( f \) is defined as the family of sets consisting of its syntactic iterative application, semantic syntactic expansion, and logical typing.  The properties of \( S(f) \) are referred to as the \textbf{Construction Defining Functionality (CDF)} of \( f \).\\

\textbf{Example 1.} Let \( f(x) := x + 1 \) over the natural numbers \( \mathbb{N} \). The structural space \( S(f) \) generated by \( f \) is defined as:

\begin{itemize}
  \item \textbf{Syntactic Component} \( S_{\mathrm{syn}}(f) \):
  \[
  S_{\mathrm{syn}}(f) := \{ f^n(0) \mid n \in \mathbb{N} \} = \{ 0, 1, 2, 3, \dots \}
  \]
  This is the orbit of 0 under repeated application of \( f \), forming a countable, linearly ordered set.

  \item \textbf{Semantic Component} \( S_{\mathrm{sem}}(f) \):
  \[
  S_{\mathrm{sem}}(f) := \left\{ \varphi(x, y) \,\middle|\, \varphi(x, y) \ :=y = x + 1,\ x, y \in \mathbb{N} \right\}
  \]
  This set represents the graph of the function \( f \), understood as a binary relation that is stable and non-branching.

  \item \textbf{Logical Component} \( S_{\mathrm{log}}(f) \):
  \[
  S_{\mathrm{log}}(f) := \left\{ T \in \mathcal{T} \,\middle|\, T \text{ is a linear computation tree without branching} \right\}
  \]
  where \( \mathcal{T} \) is the set of all computation trees. \\
\end{itemize}

Thus, the full structural space is:
\[
S(f) := \{ S_{\mathrm{syn}}(f),\ S_{\mathrm{sem}}(f),\ S_{\mathrm{log}}(f) \}
\]
which corresponds to a linearly ordered structure \( (\mathbb{N}, <) \).

\vspace{1em}

\textbf{Example 2.} Let \( f \) be a recursive syntactic tree generation function (e.g., a grammar-generating device). Then:

\begin{itemize}
  \item \textbf{Syntactic Component} \( S_{\mathrm{syn}}(f) \):
  \[
  S_{\mathrm{syn}}(f) := \left\{ t \in \mathcal{T} \,\middle|\, \text{\( t \) is a finite or infinite syntactic tree generated by recursive expansion} \right\}
  \]
  where \( \mathcal{T} \) denotes the set of term trees. These trees are generated via recursive unfolding of non-terminal symbols.

  \item \textbf{Semantic Component} \( S_{\mathrm{sem}}(f) \):
  \[
  S_{\mathrm{sem}}(f) := \left\{ \tau(t) \,\middle|\, t \in S_{\mathrm{syn}}(f),\ \tau \text{ is a type-theoretic interpretation} \right\}
  \]
  This includes semantic interpretations such as typing judgments, mapping syntactic forms to their semantic categories.

  \item \textbf{Logical Component} \( S_{\mathrm{log}}(f) \):
  \[
  S_{\mathrm{log}}(f) := \left\{ A \in \mathcal{A} \,\middle|\, A \text{ is a non-deterministic automaton modeling the generation process} \right\}
  \]
  where \( \mathcal{A} \) is the class of automata. 
\end{itemize}

Therefore, the full structural space S(f) corresponds to a branching tree structure associated with generative grammar.\\

\noindent\textbf{Remark.} 
Structural correspondences or interpretation mappings between these components can be introduced, allowing for interactions between the syntactic, semantic, and logical aspects of \( f \). \\

Some typical examples of such correspondences include:

\begin{enumerate}
  \item \textbf{Semantic Mapping:}  
   A map \(:S_{\mathrm{syn}}(f) \to S_{\mathrm{sem}}(f)\) that assigns syntactic structures to semantic interpretations, such as type assignments or meaning functions.

  \item \textbf{Reconstruction of Logical Structure:}  
  A map \(: S_{\mathrm{sem}}(f) \to S_{\mathrm{log}}(f)\) that translates semantic objects into logical or computational structures, e.g., computation trees or proof objects.

  \item \textbf{Direct Correspondence between Syntax and Logic:}  
  A map \(: S_{\mathrm{syn}}(f) \to S_{\mathrm{log}}(f)\) reflecting the embedding of syntactic constructs into logical frameworks, such as typed syntax trees corresponding to linear computation trees.
\end{enumerate}

\section{Basic Procedure of CDF framework}

This section formalizes the core concepts of the CDF framework by defining the structural space generated by a function and the interrelated components that capture its syntactic, semantic, and logical properties.

\begin{definition}[Formal Construction Procedure of CDF]\label{def:cdf-formal-construction}
Let \( f : X \to Y \) be a function and fix a basepoint \( x_0 \in X \). The Construction Defining Functionality (CDF) of \( f \), denoted by
\[
S(f) = \left( S_{\mathrm{syn}}(f), S_{\mathrm{sem}}(f), S_{\mathrm{log}}(f) \right),
\]
is generated via the following formal procedure:

\begin{enumerate}
    \item \textbf{Syntactic Component} \( S_{\mathrm{syn}}(f) \):
    \[
    S_{\mathrm{syn}}(f) := \{ f^n(x_0) \mid n \in \mathbb{N} \},
    \]
    representing the symbolic orbit generated by iterated applications of \( f \) from \( x_0 \).
    

    \item \textbf{Semantic Component} \( S_{\mathrm{sem}}(f) \):
    \[
    S_{\mathrm{sem}}(f) := \{ \phi(x,y) \mid \phi(x,y) := (y = f(x)),\ x \in X, y \in Y \},
    \]
    which captures the input-output relation graph of \( f \).


    \item \textbf{Logical Component} \( S_{\mathrm{log}}(f) \):
    \[
    S_{\mathrm{log}}(f) := \{ T(x) \in \mathcal{T} \mid T(x) \text{ is a formal computation or proof tree deriving } f(x) \},
    \]
    where \( \mathcal{T} \) denotes a suitable class of formal derivation objects.


\end{enumerate}

The intercomponent mappings are given by:
\[
\alpha_f(f^n(x_0)) := \phi(f^{n-1}(x_0), f^n(x_0)),
\]
\[
\beta_f(\phi(x,y)) := T(x \vdash y),
\]
\[
\gamma_f(f^n(x_0)) := \text{computation or derivation of } f^n(x_0).
\]


This procedure canonically lifts \( f \) into its structural analytic space \( S(f) \), embedding it into the CDF framework.
\end{definition}

\begin{remark}[On the Syntax Comprising the Syntactic Component]
The language used to construct the base syntax of the CDF structure is arbitrary and does not depend on any specific formalism. Various notations can be employed, including algebraic expressions, logical formulas, and graphical representations, all of which are mapped onto the semantic space as syntactic manifolds. A future objective is to explore a universally minimal syntactic structure that derives the semantic space using only the minimal necessary symbols, operational units, and construction rules.
\end{remark}

\begin{remark}[Note on Extensional/Intensional Terminology]
The notion of intension and extension used in CDF should not be confused with the classical set-theoretic distinction between intensional and extensional notation (e.g., $\{ x \mid P(x) \}$ vs $\{a,b,c\}$). Instead, CDF refers to the generative and observable structural dimensions of functional behavior, aligned with deeper semantic and computational layers.
\end{remark}

\begin{remark}[On Proof Trees and Computation Trees]
The logical component $S_{\mathrm{log}}(f)$ consists of formal derivation objects such as \emph{proof trees} or \emph{computation trees}. These are hierarchical structures that represent the process by which the value of $f(x)$ is logically or computationally derived.

Each node in such a tree corresponds to an inference or computation step, and the tree as a whole traces the construction of $f(x)$ from its base definitions and rules. In the case of recursive functions, these trees naturally express the nested structure of calls and their justifications, forming a concrete representation of the internal logic that governs the function's behavior.

For example, in the case of the factorial function \( f(n) = n! \), the computation of \( f(3) = 6 \) proceeds as follows:

\begin{align*}
f(3) &= 3 \cdot f(2) \\
     &= 3 \cdot (2 \cdot f(1)) \\
     &= 3 \cdot (2 \cdot (1 \cdot f(0))) \\
     &= 3 \cdot (2 \cdot (1 \cdot 1)) \\
     &= 3 \cdot 2 \cdot 1 = 6
\end{align*}

This sequence illustrates the recursive unfolding of the function's definition. Each line corresponds to a s

\end{remark}

\begin{example}[CDF of a Linear Function: \( f(x) = x + 1 \)]
Let \( f : \mathbb{N} \to \mathbb{N} \), \( f(x) = x + 1 \), and \( x_0 = 0 \). Then,
\[
S_{\mathrm{syn}}(f) = \{ n \mid n \in \mathbb{N} \},
\]
\[
S_{\mathrm{sem}}(f) = \{ \phi(n, n+1) \mid n \in \mathbb{N} \},
\]
\[
S_{\mathrm{log}}(f) := \left\{ T_n \in \mathcal{T} \;\middle|\; T_n \vdash_{\mathcal{P}} f(n) = n + 1 \right\}
\]

The mappings satisfy
\[
\alpha_f(n) = \phi(n-1,n), \quad \beta_f(\phi(n,n+1)) = T_n, \quad \gamma_f(n) = T_n,
\]
and the corresponding structural diagram commutes with stability.
\end{example}

\begin{example}[CDF of a Recursive Function: Factorial]
Consider \( f : \mathbb{N} \to \mathbb{N} \) defined by
\[
f(n) = \begin{cases}
1 & n = 0, \\
n \cdot f(n-1) & n > 0,
\end{cases}
\]
with basepoint \( x_0 = 3 \). Then, the components of the CDF for this function are:\\

1. Syntactic Component \( S_{\mathrm{syn}}(f) \):
   \[
   S_{\mathrm{syn}}(f) = \{ f(3), f(2), f(1), f(0) \},
   \]
   reflecting the recursive calls made by \( f \), such as \( f(3) \), \( f(2) \), and so on.\\

2. Semantic Component \( S_{\mathrm{sem}}(f) \):
   \[
   S_{\mathrm{sem}}(f) = \{ \phi(n, f(n)) \mid n = 0, \dots, 3 \},
   \]
   which encodes the input-output relationships of the function. For example, we have \( \phi(3, 6) \), \( \phi(2, 2) \), and \( \phi(1, 1) \), representing the relations \( f(3) = 6 \), \( f(2) = 2 \), and \( f(1) = 1 \).\\

3. Logical Component \( S_{\mathrm{log}}(f) \):
   \[
   S_{\mathrm{log}}(f) = \{ T_n \in \mathcal{T}_{\mathrm{rec}} \mid T_n \text{ is a finite tree such that } T_n \vdash_{\mathcal{R}} f(n) \},
   \]
   where \( T_n \) is the formal derivation tree showing how \( f(n) \) is computed. For example, for \( f(3) \), the corresponding computation tree \( T_3 \) shows how \( f(3) = 3 \cdot f(2) \) is derived from \( f(2) = 2 \cdot f(1) \), and so on.\\

These components are connected via the following mappings:\\

1. Syntax to Semantics Mapping (\( \alpha_f \)):
   This mapping describes how the syntactic steps (function applications) are related to the semantic input-output pairs. For example:
   \[
   \alpha_f(f^3(3)) = \phi(2, 6), \quad \alpha_f(f^2(3)) = \phi(1, 2), \quad \alpha_f(f^1(3)) = \phi(0, 1).
   \]\\

2. Semantics to Logic Mapping (\( \beta_f \)):
   The semantic relations \( \phi(x, y) \) are mapped to formal computation trees \( T(x \vdash y) \). For example:
   \[
   \beta_f(\phi(3, 6)) = T_3, \quad \beta_f(\phi(2, 2)) = T_2, \quad \beta_f(\phi(1, 1)) = T_1.
   \]
   Here, \( T_3 \), \( T_2 \), and \( T_1 \) are the formal derivation trees showing how the values \( f(3) = 6 \), \( f(2) = 2 \), and \( f(1) = 1 \) are computed.\\

3. Syntax to Logic Mapping (\( \gamma_f \)):
   The syntax is mapped directly to the derivation or computation tree. For example:
   \[
   \gamma_f(f^3(3)) = T_3, \quad \gamma_f(f^2(3)) = T_2, \quad \gamma_f(f^1(3)) = T_1.
   \]\\
   This mapping shows the derivation steps for each function application.

Thus, the mappings are defined analogously, ensuring that the structure of the recursive function's syntactic, semantic, and logical components is stable when the input is fixed.

\end{example}

\begin{remark}[Implications of Input Variability and Its Connection to NP Problems]
The structure and behavior of recursive functions become notably more complex as the input varies. In particular, when the input is not fixed, the syntactic, semantic, and logical mappings may no longer remain stable. This reflects the dynamic nature of recursive computations, where the depth of recursion, the number of calls, and the resulting computation tree can grow exponentially with respect to the input size.

\begin{itemize}
    \item \textbf{Input Variability and Instability:} 
    For recursive functions, changes in the input lead to significant variations in the computation process. In cases like the factorial function, larger inputs result in deeper recursive calls, which cause a dramatic increase in the complexity of the associated syntactic, semantic, and logical structures. These structures become unstable when input size grows, as the function's behavior can no longer be predicted solely from the initial fixed values.

    \item \textbf{Connection to NP Problems:} 
    The behavior of recursive functions under changing inputs bears a striking resemblance to the structure of NP problems. Many NP-complete problems, such as the traveling salesman problem or 3-SAT, exhibit exponential growth in complexity as the size of the input increases. This results in vast search spaces, similar to the deep recursive trees in the factorial example. Therefore, the instability of recursive mappings when inputs change mirrors the computational hardness observed in NP problems, where the solution space expands exponentially and requires complex exploration.
\end{itemize}

Thus, the interplay between recursive functions and NP problems emphasizes how the growth of computational complexity, driven by input variability, can lead to unmanageable computation and logical structures. This connection serves as an important bridge for understanding the deeper computational challenges faced in both theoretical and practical problem-solving contexts.
\end{remark}

\section{Theoretical Perspectives on Function Complexity via CDF}

This section discusses two key theoretical perspectives that contribute to the foundation of Construction Defining Functionality (CDF). Here, we aim to position CDF as a framework that classifies and articulates the structural and logical complexity of functions.  The first theme examines the tree-like nature of theories and how it reflects the structural behavior of functions.The second theme explores the relation between functions and logical expressibility.

\subsection{Tree Structure and the Complexity of Theories}

Many logical theories can be distinguished based on whether they exhibit tree-like branching structures. Such structural features directly affect the nature of functions definable within the theory and the behavior of their generated structural spaces in the sense of CDF.

\begin{table}[h]
\centering
\small
\caption{Comparison: Tree-Structured vs. Non-Tree-Structured Theories}
\begin{tabularx}{\textwidth}{|l|X|X|}
\hline
\textbf{Feature} & \textbf{Tree-Structured} & \textbf{Non-Tree-Structured} \\
\hline
Structural complexity & High (recursive, branching) & Limited (linear, simple) \\
Classifiability & Unclassifiable & Classifiable \\
Examples & Peano Arithmetic (PA), ZFC & Theory of Abelian Groups, Real Closed Fields \\
Model diversity & Often uncountable models & More uniform, tame models \\
\hline
\end{tabularx}
\end{table}

Tree-structured theories tend to:

\begin{itemize}
  \item Allow complex choices and extensions of substructures.
  \item Lose desirable model-theoretic properties such as stability, simplicity, or NIP.
  \item Exhibit non-constructive or non-deterministic traits in proof theory and computability.
\end{itemize}

These behaviors correspond closely to the complexity of the logical component $S_{\mathrm{log}}(f)$ and syntactic expansion $S_{\mathrm{syn}}(f)$ in the CDF of a given function $f$.

\paragraph{Model-Theoretic Properties and Tree-Likeness}

\begin{table}[h]
\centering
\caption{Key Properties and Their Relationship to Tree-Likeness}
\begin{tabular}{|l|p{5.0cm}|p{6.0cm}|}
\hline
\textbf{Property} & \textbf{Meaning} & \textbf{Relation to Tree Structure} \\
\hline
Stability & Bounded number of types & Implies absence of tree-likeness \\
Simplicity & Forking is well-behaved & Requires absence of TP$_2$ \\
NIP & No independence patterns & Indicates limited branching (no IP) \\
Classifiability & Models are tame and well-ordered & Presupposes non-tree-like structure \\
\hline
\end{tabular}
\end{table}

In summary, a function whose CDF exhibits deep branching or recursive unfolding tends to correspond to theories that are unclassifiable or unstable. Conversely, linearly structured functions (e.g., arithmetic progression) lead to tame, classifiable theories.

\subsection{Logical Expressibility and Function Classes}

The complexity of a function $f : X \to Y$ can also be captured by the logical strength required to define or describe it. CDF enables us to analyze functions based on the minimal logical framework in which their structure is expressible.

\begin{table}[h]
\centering
\caption{Levels of Logical Expressibility for Functions}
\begin{tabular}{|l|p{5.3cm}|p{5.5cm}|}
\hline
\textbf{Logic} & \textbf{Expressible Functions} & \textbf{Example} \\
\hline
First-order logic (FOL) & Functions with explicit algebraic definition & $f(x, y) = x + y$ \\
Higher-order logic (HOL) & Functions‐over‐functions (Lambda-abstractions) & Typed Lambda-calculus \\
Recursive function theory & Computable, (partial) recursive functions & Ackermann‐function,halting predicates \\
Type theory & Constructive, proof-relevant functions & Curry-Howard‐correspondence \\
\hline
\end{tabular}
\end{table}

In the CDF framework:
\begin{itemize}
  \item $S_{\mathrm{syn}}(f)$ reflects the syntactic depth or recursive structure.
  \item $S_{\mathrm{sem}}(f)$ captures the semantic interpretation in a given logic.
  \item $S_{\mathrm{log}}(f)$ encodes the logical or computational mechanisms for realization.
\end{itemize}

Thus, the classification of a function is not only based on its domain and codomain, but also on the complexity of the logic necessary to meaningfully express and analyze it.

\paragraph{Summary}

CDF‐theory provides a structural view on functions beyond their operational definition. It enables classification according to:
\begin{itemize}
  \item Logical definability: What kind of logic suffices to express $f$?
  \item Structural generation: How complex is the expansion of $f$ syntactically and semantically?
  \item Theoretical impact: What type of theory does $f$ induce or correspond to (stable, classifiable, etc.)?
\end{itemize}

By situating functions within these theoretical spectra, CDF becomes a tool not just for analyzing individual functions, but for describing the structure of entire theories they may define or inhabit.\\

\section{Classification Framework of Functions via Construction Defining Functionality (CDF)}

In this section, we present a foundational classification framework for functions based on their structural and logical complexity within the Construction Defining Functionality (CDF) theory.  
The classification is organized into a hierarchical scheme reflecting the complexity of the generated structures such as models, tree-like branching, and logical formulas.  
Additionally, supplementary classifications capture other important functional characteristics that do not fit neatly into the hierarchy.

\subsection{Hierarchical Classification}

The core framework classifies functions into levels according to the complexity of their generated structures:

\begin{table}[h]
\centering
\caption{Hierarchical Classification of Functions by CDF}
\renewcommand{\arraystretch}{1.3} 
\begin{tabular}{|c|p{3.0cm}|p{5.0cm}|p{5.0cm}|}
\hline
\textbf{Level} & \textbf{Class} & \textbf{Description} & \textbf{Examples} \\
\hline
0 & Linear Function Class & Functions generating no tree structure, only linear or ordered structures & \( f(x) = x + 1 \), linear mappings \\
\hline
1 & Finite‐Branching Structure Functions & Functions generating finite branching structures & Grammar generation functions, finite automata \\
\hline
2 & Infinite‐Tree Structure‐Functions & Functions generating infinite, recursive tree structures & Recursive grammars, proof tree generators \\
\hline
3 & Functions with Logical‐Tree Properties & Functions encoding logical tree properties such as TP$_1$, TP$_2$ & Functions corresponding to complex logical theories \\
\hline
4 & Higher-Order Tree Structure‐Functions & Functions encoding higher-order, set-theoretic structures & Functions that generate and manipulate hierarchical set-theoretic structures defined in ZFC
\\
\hline
\end{tabular}
\end{table}

\vspace{1em}
\noindent
The classification table above organizes function classes by the complexity of the structures they generate. We briefly explain each level below to clarify their theoretical implications.

\begin{itemize}
  \item \textbf{Level 0 (Linear Function Class):} These functions generate no branching or hierarchical structure. Their outputs are simple sequences or ordered chains, corresponding to basic operations like arithmetic progression or simple iteration.
  
  \item \textbf{Level 1 (Finite‐Branching Structure Functions):} These functions introduce discrete, bounded branching. They often appear in the generation of syntactic trees with limited depth, such as those from regular or context-free grammars, and are related to finite automata or simple parsers.
  
  \item \textbf{Level 2 (Infinite‐Tree Structure Functions):} These functions define or explore infinite recursive structures. Examples include derivation trees in proof systems, recursive grammars, or abstract syntax trees produced by recursive descent algorithms. The trees may be countably infinite and often involve self-reference.
  
  \item \textbf{Level 3 (Functions with Logical‐Tree Properties):} At this level, functions produce trees with advanced logical properties such as the tree property TP$_1$ or TP$_2$, which arise in model theory and infinitary logic. These trees reflect combinatorial and logical constraints that go beyond first-order definability, and the functions are associated with classification theory or stability analysis in logic.
  
\item \textbf{Level 4 (Higher-Order Tree Structure Functions):} Functions at this highest level manipulate or define higher-order structures, including trees whose nodes themselves may be entire trees or sets. These functions are relevant to second-order logic, type theory, and particularly to set theory. 

\end{itemize}

  In the context of ZFC (Zermelo-Fraenkel set theory with Choice), such functions can define the \emph{cumulative hierarchy} \( V_\alpha \), a transfinite construction where each level \( V_{\alpha+1} \) is the powerset of \( V_\alpha \), and limit stages aggregate all earlier levels. This hierarchy models the universe of all sets and grows through the ordinals. Functions at this level may construct or traverse such hierarchies, define large cardinals, or operate over inner models and forcing extensions. These operations reflect a level of structural complexity that exceeds computability or definability in first-order logic.

This hierarchy captures the increasing complexity from simple linear functions up to highly complex functions corresponding to advanced logical and set-theoretic frameworks.

\subsection{Supplementary Classification}

In addition to the hierarchical classification, certain function classes exhibit characteristics orthogonal to structural complexity, necessitating separate treatment:

\begin{table}[h]
\centering
\caption{Supplementary Function Classes in CDF}
\renewcommand{\arraystretch}{1.3}

\begin{tabular}{
    |>{\raggedright\arraybackslash}p{3.5cm}
    |>{\raggedright\arraybackslash}p{6.2cm}
    |>{\raggedright\arraybackslash}p{5.0cm}|
}
\hline
\textbf{Class} & \textbf{Description} & \textbf{Examples} \\
\hline
Periodic and Continuous Functions & Functions exhibiting periodicity, continuity, or analytic properties not captured by tree structure & Trigonometric functions (\(\sin, \cos\)), exponential functions \\
\hline
Probabilistic‐Random Functions & Functions whose outputs are governed by probabilistic distributions or randomness & Stochastic processes, Markov chains \\
\hline
Nondeterministic‐Concurrent Functions & Functions incorporating nondeterminism or parallel execution semantics & Nondeterministic automata, concurrent computation models \\
\hline
Generalized Functions and Distributions & Extensions of classical functions, including generalized functions used in analysis & Delta distributions, Green's functions \\
\hline
Quantum‐Functions & Functions defined in the quantum computing framework, involving linear algebraic operators & Quantum gates, unitary transformations \\
\hline
Non-measurable/Non-constructive Functions & Functions that defy constructive or measurable description & Functions related to the Banach-Tarski paradox \\
\hline
\end{tabular}
\end{table}

These supplementary classes reflect important functional behaviors and theoretical concepts that complement the hierarchical classification.

\subsection{Remarks}

The CDF-based classification framework thus provides a structured, extensible approach to analyzing functions according to both their generative structural complexity and additional functional properties.  
It enables a nuanced understanding of how functions relate to logical theories, computational models, and analytic properties.  
Future work may focus on exploring mappings and correspondences between classes, as well as expanding the framework to accommodate emerging classes of functions.

\section{Classification of Function-Generated Structural Spaces}

This section introduces a classification framework for the structural space \( S(f) \) generated by a function \( f \). Central to the CDF perspective is the idea that a function should not be understood solely by its input-output behavior, but by the \textit{structural space} it generates through syntactic unfolding, semantic interpretation, and logical computation.

We classify \( S(f) \) using three \textbf{core axes} and two \textbf{derivative axes}, each reflecting a different aspect of the space’s geometry, dynamics, and logic.

\subsection*{A. Structural Shape (Geometry / Topology)}

This axis categorizes the geometric or graph-theoretic form of the structure generated by the function. It is the most fundamental, as it directly impacts all other properties.

\begin{itemize}
  \item \textbf{Linear structures}: e.g., $f(x) = x + 1$ produces a sequence or line.
  \item \textbf{Cyclic structures}: e.g., $f(x) = \sin(x)$ generates a periodic orbit.
  \item \textbf{Branching structures}: such as trees from recursive or grammar-based functions.
  \item \textbf{Graph/network structures}: general graphs from non-linear, interconnected functions.
  \item \textbf{Layered/hierarchical structures}: e.g., category-like or stratified definitions.
\end{itemize}

\subsection*{B. Dynamic Expandability (Recursion / Behavior)}

This axis examines how the structure evolves under iteration of the function. It captures recursive depth, periodicity, and potential divergence—concepts closely related to dynamical systems.

\begin{itemize}
  \item \textbf{Closed/finite behavior}: producing bounded or periodic structures.
  \item \textbf{Recursive expansion}: inductive or recursively unfolding patterns.
  \item \textbf{Infinite branching}: e.g., recursive proof trees or unbounded generation.
  \item \textbf{Chaotic/divergent behavior}: such as in the tent map or logistic map.
\end{itemize}

\subsection*{C. Logical Complexity (Model-Theoretic Classification)}

This axis captures how complex the generated structure is in model-theoretic terms, focusing on properties such as stability, simplicity, independence, and tree properties.

\begin{itemize}
  \item \textbf{Stable}: bounded number of types.
  \item \textbf{Simple}: controlled forking and predictable type structure.
  \item \textbf{NIP}: no independence property (e.g., o-minimal structures).
  \item \textbf{TP\textsubscript{1}/TP\textsubscript{2}}: tree-like logical behavior.
  \item \textbf{Non-classifiable}: chaotic or highly branched theories.
\end{itemize}

\subsection*{D. Computability-Theoretic Properties (Derived Axis)}

A supplemental axis focusing on algorithmic aspects—whether the function or its generated structure can be effectively computed or approximated.

\begin{itemize}
  \item \textbf{Computable vs Non-computable}
  \item \textbf{Recursive vs Non-recursive}
  \item \textbf{Turing Halting Behavior}
  \item \textbf{Complexity Classes}: P, NP, EXP, etc.
\end{itemize}

\subsection*{E. Descriptive and Analytic Properties (Derived Axis)}

This axis covers analytic and descriptive properties such as continuity, differentiability, or algebraicity. These reflect the smoothness or definability of the function's behavior.

\begin{itemize}
  \item \textbf{Injectivity, Surjectivity, Bijectivity}
  \item \textbf{Continuity, Differentiability}
  \item \textbf{Integrability, Analyticity}
  \item \textbf{Algebraic vs Transcendental}
\end{itemize}

\subsection{Summary: Interpretive Roles of the Axes}

In summary, the structure space \( S(f) \) generated by a function \( f \) can be understood as arising from the interplay of three fundamental axes:

\begin{itemize}
  \item \textbf{A: Structural Shape} — representing the topological or combinatorial characteristics of the generated space.
  \item \textbf{B: Expandability} — describing how the structure evolves or grows under iterative application of the function.
  \item \textbf{C: Logical Complexity} — capturing the logical sophistication required to define, classify, or reason about the structure.
\end{itemize}

Rather than acting independently, these axes combine in a nuanced way to produce a rich spectrum of structural complexity. The emergent complexity of the generated space reflects this multidimensional interaction.

Other axes, such as computability (D) and analytic properties (E), are often derivative or subordinate to these core dimensions, providing additional but secondary perspectives.

This multifaceted framework offers a robust foundation for classifying and comparing functions based on the complexity of the structures they generate.

\section{Master Table of Function-Generated Structural Classifications}

To consolidate the classification system, we present the following master table that organizes structural spaces \( S(f) \) according to the five key axes. The first three axes (A–C) form the core of the theory, while D and E serve as complementary dimensions that provide computational and descriptive refinement.

\begin{longtable}{@{}p{0.35\textwidth} p{0.65\textwidth}@{}}
\toprule
\textbf{Axis of Classification} & \textbf{Types of Structure} \\
\midrule
\textbf{A.Geometric/Topological Shape} &
- Linear (sequences, number lines) \newline
- Cyclic (circles, tori) \newline
- Branching (trees, forests) \newline
- Network-like (general graphs) \newline
- Layered / Hierarchical (categories, stratified structures) \\
\textbf{B.Dynamical‐Expandability} &
- Closed (finite, periodic) \newline
- Iterative expansion (recursion, induction) \newline
- Infinite branching (recursive trees) \newline
- Divergent / chaotic (non-periodic, unbounded choice) \\
\textbf{C.Logical Complexity} &
- Stable (bounded number of types) \newline
- Simple (controlled forking structure) \newline
- NIP (no independence property) \newline
- TP\textsubscript{1}/TP\textsubscript{2} (tree properties present) \newline
- Non-classifiable (chaotic, unpredictable) \\
\textbf{D.Computability Properties} &
- Computable vs non-computable \newline
- Recursive vs non-recursive \newline
- Turing halting status \newline
- Complexity classes (P, NP, EXP, etc.) \\
\textbf{E.Descriptive/Analytic Properties} &
- Injectivity, surjectivity, bijectivity \newline
- Continuity, differentiability \newline
- Integrability, analyticity \newline
- Algebraic vs transcendental \\
\bottomrule
\end{longtable}

The table above functions as a conceptual taxonomy for structural spaces induced by functions. Each axis corresponds to a theoretical lens, and their intersections allow for multidimensional classification. In particular, core axes A–C establish the ontological structure of \( S(f) \), while D and E offer further specification in computational and analytic terms.

\section{Analytical Methods by Evaluation Criteria}

To support practical analysis and comparison, the following table presents concrete evaluation metrics, methods, and examples associated with each classification axis.

\begin{longtable}{
  >{\raggedright\arraybackslash}p{0.19\textwidth}
  >{\raggedright\arraybackslash}p{0.16\textwidth}
  >{\raggedright\arraybackslash}p{0.17\textwidth}
  >{\raggedright\arraybackslash}p{0.19\textwidth}
  >{\raggedright\arraybackslash}p{0.17\textwidth}
}
\toprule
\textbf{Classification Axis} & \textbf{Evaluation Metric} & \textbf{Methodology} & \textbf{Criteria} & \textbf{Examples} \\
\midrule
\endfirsthead
\toprule
\textbf{Classification Axis} & \textbf{Evaluation Metric} & \textbf{Methodology} & \textbf{Criteria} & \textbf{Examples} \\
\midrule
\endhead

\textbf{A.Structural Shape} & Geometry & Graph-based classification & Cycles, degree, branching & Term-trees, graph theory \\
& Branching degree & Tree-depth / branching factor & Max outdegree & Recursive trees, syntax trees \\
\midrule
\textbf{B.Expandability} & Periodicity & Orbit analysis & Periodic points, cycle length & Modular maps, dynamical systems \\
& Recursivity & Recursive function iteration & Iterative behavior & $f(n) = 2n$, etc. \\
& Chaoticity & Sensitivity, divergence analysis & Lyapunov exponents & Logistic map, tent map \\
\midrule
\textbf{C.Logical Complexity} & Stability & Comparison of type spaces & Cardinality bounds & Stability theory (e.g., PA vs Q) \\
& Simplicity & Forking behavior analysis & Dividing controlled & Simple theories \\
& NIP & Evaluation of independence & Absence of IP & O-minimal theories \\
& Tree‐Property (TP) & Existence of infinite branches & TP\textsubscript{1}/TP\textsubscript{2} criteria & Classification theory \\
\midrule
\textbf{D.Computability} & Recursiveness & Turing computability analysis & Computable or not & Partial recursive functions \\
& Halting behavior & Reduction to halting problem & Oracle dependence & Turing undecidable cases \\
& Complexity & Time/space complexity & Class membership & SAT, NP-complete functions \\
\midrule
\textbf{E.Descriptive Properties} & Continuity & $\varepsilon$-$\delta$ definition & Pointwise continuity & Step functions, continuous maps \\
& Differentiability & Existence of derivative & Domain of differentiability & $f(x)=|x|$, etc. \\
& Integrability & Integral value existence & Riemann/Lebesgue integrability & Indicator functions \\
& Algebraicity & Defined by polynomials & Satisfiability of equations & $e^x$, $\pi$: transcendental \\
\bottomrule
\end{longtable}

This table provides concrete tools for analyzing function-generated spaces across axes. By specifying both theoretical criteria and practical methods, it enables comparative study and empirical classification within the CDF framework.

\section{Future Directions and Applications}

Building on this classification framework, the following potential developments can be considered:

\begin{itemize}
  \item Viewing functions as ``structure-generating devices,'' enabling theoretical comparison of the properties of the structures they produce. \\
    \textit{For example, quantitative evaluation of functions using CDF, such as defining a structural generation capacity index.}
  \item Applying the classification of functional structural complexity hierarchies in fields such as AI, natural language processing, and automated theorem proving.
  \item Exploring the potential of CDF as an integrated theory unifying logic, syntax, computation, and spatial structures.
  \item Establishing a unified language that connects existing theories, including dynamical systems, model theory, and topology.
\end{itemize}

Overall, this framework provides a useful foundation for understanding the complexity of structure-generating functions. While further refinement and validation are needed, it offers a direction toward bridging concepts in logic, computation, and spatial structures, with potential applications across various scientific and engineering fields.

\nocite{*}
\bibliographystyle{plain}
\bibliography{references}

\end{document}